\documentclass[conference]{IEEEtran}
\IEEEoverridecommandlockouts
\usepackage{cite}
\usepackage{amsmath,amssymb}
\usepackage{graphicx}
\usepackage{booktabs}
\usepackage{xcolor}
\usepackage[hidelinks]{hyperref}
\usepackage{tabularx}
\usepackage{colortbl}
\usepackage{graphicx}
\usepackage[caption=false,font=footnotesize]{subfig}

\definecolor{LaneDropTint}{HTML}{EAF2FC}

\begin{document}

\title{World-Model-Guided Plan Search for Cooperative Mixed-Traffic Control at Bottlenecks}

\author{
\IEEEauthorblockN{Jinghan Peng\textsuperscript{1}, Ning Liu\textsuperscript{2}, Xiangrui Kong\textsuperscript{2}, Sen Shen\textsuperscript{3}, Dongkun Han\textsuperscript{3}*}
\IEEEauthorblockA{\textsuperscript{1}Department of Electrical and Computer Engineering, \\
University of California, Los Angeles, CA, USA} 
\IEEEauthorblockA{\textsuperscript{2}Department of Electrical, Electronic and Computer Engineering, School of Engineering,\\
The University of Western Australia, Perth, WA, Australia} 
\IEEEauthorblockA{\textsuperscript{3}Department of Mechanical and Automation Engineering, The Chinese University of Hong Kong, Hong Kong SAR, China} 
\IEEEauthorblockA{*Corresponding author: Dongkun Han (dkhan@mae.cuhk.edu.hk)}
}

\maketitle

\begin{abstract}
Cooperative speed control by connected and automated vehicles can reduce congestion at lane drop bottlenecks. However, effective coordination depends on vehicle roles and execution timing, while initial holding costs may be offset by later traffic benefits. This paper proposes an offline plan search framework that combines world model guidance with simulator verification. A traffic world model incorporating physical dynamics guides evolutionary search over vehicle roles, target speeds, holding durations, and start times. Paired SUMO simulations screen candidates for total delay reduction and limits on additional mean delay across vehicle groups. Accepted plans form a reusable library whose actions temporarily override a frozen MAPPO policy. Across 192 paired test scenarios at 60\% penetration and a demand of 2,800 vehicles per hour, the searched library reduces mean total delay by 14.3\% relative to MAPPO and by 3.8\% relative to a handwritten plan library. The largest increase in group mean delay is 6.2~s, below the prescribed 10~s limit. In a separate screening experiment, the guided search achieves a 30.0\% pass rate compared with 2.6\% for random sampling. These results support the use of predictive search and independent verification for discovering cooperative plans under group delay constraints.
\end{abstract}

\begin{IEEEkeywords}
Connected automated vehicles, mixed traffic, lane drop, cooperative control, world models, CAV penetration.
\end{IEEEkeywords}

\section{Introduction}

Lane reductions create traffic bottlenecks where vehicles from a closing lane must merge into the remaining lanes~\cite{chung2024gap}. In mixed traffic, connected and automated vehicles (CAVs), can influence surrounding traffic through coordinated speed control~\cite{lee2025field}. Controlling a small share of vehicles can improve traffic performance beyond the controlled vehicles themselves~\cite{wu2022flow}. This motivates cooperative control that reduces delay across the traffic stream rather than optimizing the progress of individual CAVs alone.

\begin{figure}[!t]
    \centering
    \includegraphics[width=\linewidth]{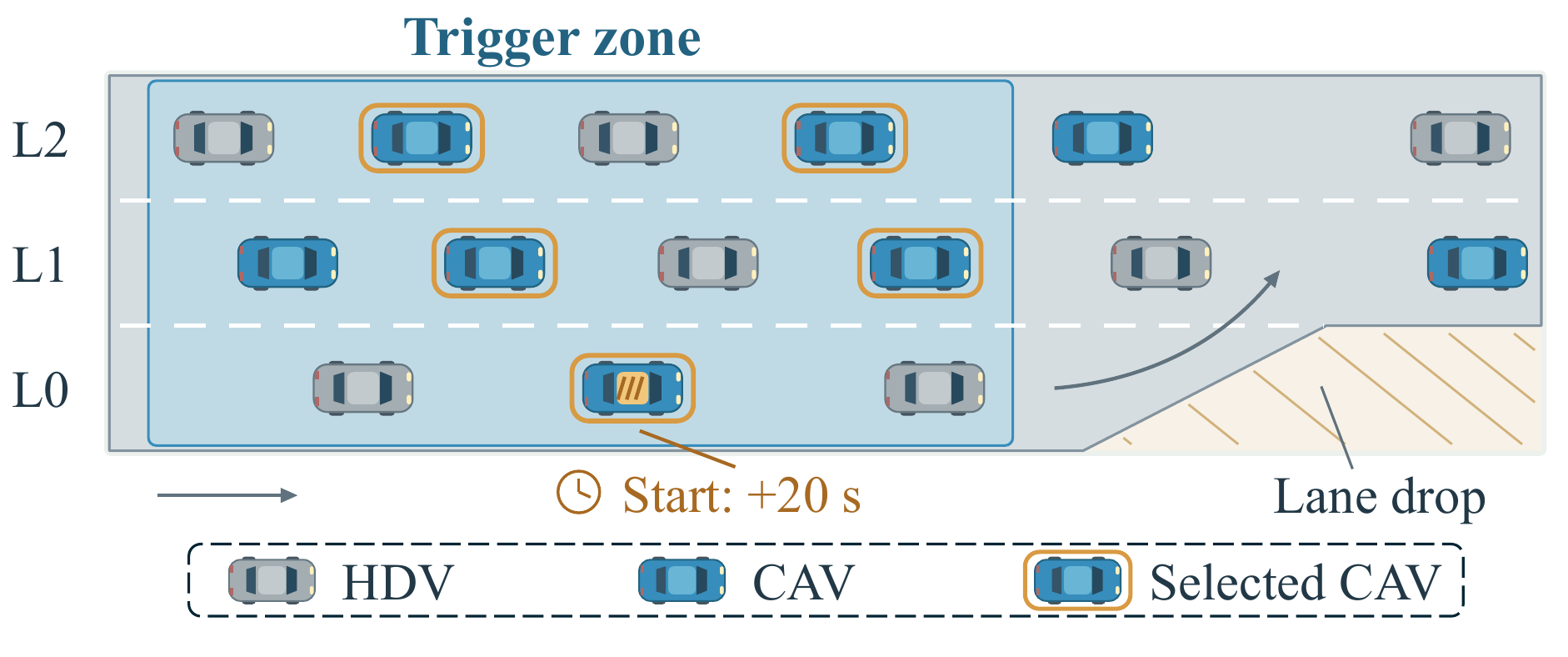}
    \caption{An example cooperative plan at a lane-drop bottleneck, involving five CAVs across three lanes. The striped CAV starts its speed-control action 20 s after the plan is triggered.}
    \label{fig:scenario}
\end{figure}

Effective cooperation depends on both vehicle roles and execution timing, as illustrated in Fig.~\ref{fig:scenario}. Coordinated trajectories couple vehicle positions and speed adjustments to facilitate merging~\cite{liu2024multistep}. A sustained control plan must specify participating vehicles, lane roles, target speeds, holding durations, and starting times. Its value depends on whether subsequent traffic benefits offset the delay introduced by holding vehicles upstream~\cite{cicic2022platoons}. These choices should therefore be considered jointly, with plans evaluated through cumulative traffic delay and the additional costs borne by different vehicle groups.

Existing methods obtain control behaviors through planning, policy learning, and combinations of the two. Unified reward design expresses driving objectives across multiple scenarios~\cite{han2024multi}. In related robotic navigation, shared exploration maps support the integration of global search and multiagent reinforcement learning~\cite{liu2025cooperative}. Hybrid exploration couples task allocation with local navigability and switches between global planning and reactive control~\cite{liu2026vorl}. Multiagent world models predict interaction trajectories to support policy learning~\cite{zhang_marie2024}. This predictive capability also motivates comparing candidate cooperative plans before committing to detailed simulation.

Discovering sustained cooperative plans requires joint exploration of vehicle roles and execution timing under a limited simulation budget. Repeated evaluation across stochastic traffic realizations makes this search computationally demanding. Surrogate models have been used to guide candidate evaluation in traffic signal optimization and coordinated ramp metering~\cite{gong2025optimizing}. However, predicted gains alone do not establish control effectiveness~\cite{oefinger_validate2026}. Effective plan discovery therefore requires predictive search to be coupled with independent verification of both cumulative traffic benefits and the additional delay borne by different vehicle groups.

To address this challenge, this paper proposes a framework for cooperative plan search guided by a physics informed world model at lane drop bottlenecks. The model guides evolutionary search over vehicle roles, target speeds, holding durations, and onset times. Shortlisted plans undergo paired SUMO evaluations against a frozen MAPPO controller and enter a reusable library only after meeting the total delay reduction and group mean delay criteria. Search and verification take place offline, while a fixed executor activates eligible plans and returns control to MAPPO after release. Across 192 paired test scenarios at 60\% CAV penetration, the resulting library reduces mean total delay by 14.3\% relative to MAPPO and by 3.8\% relative to the handwritten plan library, while keeping the additional mean delay for every evaluated vehicle group below 10~s.

Our contributions are:
\begin{itemize}
    \item A cooperative plan-search framework that jointly selects vehicle roles and execution timing, combining world-model guidance with simulator verification of traffic benefits and vehicle-group delay costs.
    \item A controlled evaluation protocol that compares model-guided and unguided candidate selection under matched simulation budgets to assess the world model's contribution to discovering beneficial plans.
    \item An evaluation design combining traffic performance measurements with resource-matched and synchronized-start comparisons to distinguish the value of cooperation structure from the effects of controlling more vehicles.
\end{itemize}

\section{Related Work}

\subsection{Cooperative Control in Mixed Traffic}
Existing studies have explored cooperative traffic control through reinforcement learning and coordinated vehicle planning. TD3 learns continuous CAV control actions to improve traffic efficiency around moving bottlenecks~\cite{lin2025moving}. Cooperative CAV fleets combine reinforcement learning for speed and spacing decisions with model predictive control for trajectory tracking~\cite{yang2026fleets}. In merging areas, separate reinforcement learning modules coordinate merging and lane changing decisions~\cite{yang2025dual}. Predictive behavior assessment further improves cooperative decisions by screening unsafe exploratory actions before execution~\cite{wei2026pretrial}. Unlike these approaches, our work focuses on jointly searching vehicle roles and execution timing, followed by explicit verification of long term traffic benefits and vehicle group delay costs.

\subsection{World Models and Simulation-Guided Search}

World models support planning by predicting how an environment evolves under different actions. TD-MPC2 uses learned latent dynamics and value estimates to optimize action sequences~\cite{hansen2024tdmpc2}. In traffic control, TrafficWise predicts traffic evolution under different signal strategies to support policy optimization~\cite{hu2025trafficwise}, while physics-informed world models generate imagined trajectories with uncertainty control for offline cooperative learning~\cite{liu2026offlinepiwm}. Model predictions can also reduce the cost of simulation-based search: surrogate ensembles have been used to identify traffic signal settings under limited simulation budgets~\cite{tang2025ade}. Building on these predictive approaches, the present work focuses on cooperative plans that specify CAV roles, speed controls, and execution timing. A traffic world model predicts how each plan influences surrounding traffic and guides the search for promising candidates. The selected plans are then evaluated in SUMO against a MAPPO baseline, considering both total-delay savings and the additional delay borne by different vehicle groups.

\section{Problem Formulation}
\label{sec:problem}

\subsection{Traffic Setting and Cooperative Plans}

The traffic setting is a three-lane freeway bottleneck in which the on-ramp lane L0 ends and vehicles must merge into the adjacent mainline lane L1. Lane L2 is the other mainline lane. Connected and automated vehicles (CAVs) follow a pretrained multi-agent proximal policy optimization (MAPPO) policy, denoted by $\pi_0$, as the baseline controller. Human-driven vehicles follow their original driving models. During the execution of a cooperative plan, a cooperative controller temporarily overrides the longitudinal control of selected CAVs without directly controlling lane changes.

A cooperative intervention depends on which CAVs participate, how they adjust their speeds, and when their control actions are applied. A cooperative plan is therefore represented by the participating vehicle roles, their speed-control parameters, and their execution timing:

\begin{equation}
\theta =
\left(
\mathcal{Z}, m,
\{r_i, v_i^{\mathrm{h}}, d_i, T_i, \tau_i\}_{i=1}^{m}
\right),
\label{eq:plan}
\end{equation}
where $\mathcal{Z}$ is the trigger zone and $m$ is the number of participating CAVs. For the $i$-th participant, $r_i$ specifies its lane role, $v_i^{\mathrm{h}}$ is its target holding speed, and $d_i$ is its maximum allowed deceleration. The parameter $\tau_i$ specifies the delay between the plan trigger time and the start of its speed-control action, while $T_i$ specifies the planned holding duration. This parameterization allows the subsequent search to compare different combinations of vehicle roles, speed adjustments, and execution times.

The verified plans form a library $\mathcal{B} = \{\theta_1,\ldots,\theta_M\}$. The controller $\pi_{\mathcal{B}}$ selects a plan when its trigger and execution conditions are satisfied. During plan execution, the selected CAVs follow their assigned speed-control actions, while CAVs not under plan control follow the baseline policy $\pi_0$. A selected CAV returns to $\pi_0$ after it is released from the plan.

\begin{figure*}
    \centering
    \includegraphics[width=1.0\textwidth]{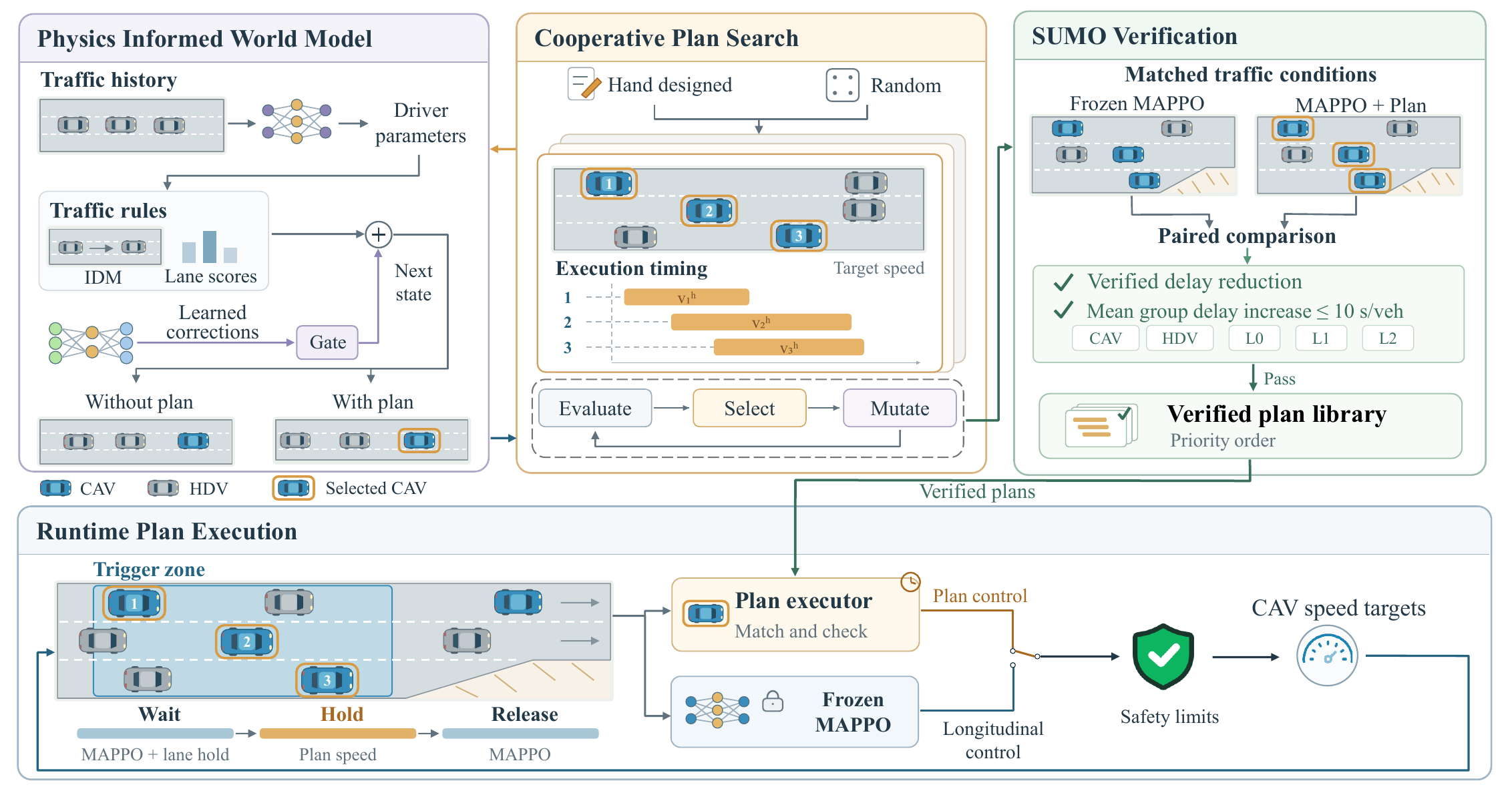}
    \caption{Overview of the proposed framework. A physics informed world model guides offline plan search. SUMO checks delay savings and group delay constraints, and accepted plans form a library for runtime execution alongside frozen MAPPO.}
    \label{fig:framework}
\end{figure*}

\subsection{Objective and Constraints}

The objective is to reduce the cumulative delay experienced by traffic at the bottleneck, including both vehicles on the road and vehicles waiting to enter. For a traffic realization $\xi$, the delay under controller $\pi$ is defined as
\begin{equation}
J(\pi,\xi)=
\sum_{t\in\mathcal T}
\left[
\sum_{v\in\mathcal V_t}
\max\left(0,1-\frac{v_{v,t}}{v_{\lim}}\right)
+N_t^{\mathrm{wait}}
\right]\Delta t,
\label{eq:delay}
\end{equation}
where $\mathcal T$ is the evaluation horizon, $\mathcal V_t$ is the set of vehicles on the road at time $t$, $v_{v,t}$ is the speed of vehicle $v$, $v_{\lim}$ is the speed limit, and $N_t^{\mathrm{wait}}$ is the number of vehicles waiting to enter the road. The first term accumulates delay associated with travel below the speed limit, while the second accounts for entrance waiting. Including both terms captures delay experienced by vehicles that have not yet entered the road.

Reducing total delay alone does not prevent a control plan from imposing additional delay on particular groups of vehicles. We therefore consider a set of vehicle groups $\mathcal G$ defined by their initial lanes and driving classes. Let $\Delta D_g(\mathcal B,\xi)$ denote the change in mean per-vehicle delay of group $g$ under $\pi_{\mathcal B}$ relative to the baseline policy $\pi_0$. We seek a plan library that reduces expected total delay while limiting the expected additional delay for each group:
\begin{subequations}
\label{eq:problem}
\begin{align}
    \min_{\mathcal{B}} \quad
    & \mathbb{E}_{\xi}\!\left[J(\pi_{\mathcal{B}},\xi)\right]
    \label{eq:problem-objective} \\
    \mathrm{s.t.} \quad
    & \mathbb{E}_{\xi}\!\left[
        \Delta D_g(\mathcal{B},\xi)
      \right]
      \leq \epsilon,
      \quad \forall g\in\mathcal G,
    \label{eq:problem-group} \\
    & N_{\mathrm{verify}}\leq B.
    \label{eq:problem-budget}
\end{align}
\end{subequations}
The expectations are taken over traffic realizations. The threshold $\epsilon$ bounds the expected increase in mean delay for each group, and $B$ limits the number of candidates that can be sent for high-fidelity simulation verification.

In the proposed approach, a world model guides the search for candidate plans under this verification budget. Individual candidates are then screened using finite-sample simulator evaluations, and the plan library is evaluated separately to assess its traffic performance and group delay impacts.

\section{Methodology}
\label{sec:method}

As shown in Fig.~\ref{fig:framework}, a world model guides offline cooperative plan search, and SUMO simulations verify the shortlisted candidates. At runtime, a fixed executor applies plans from the  library $\mathcal B$ alongside the frozen MAPPO policy $\pi_0$.

\subsection{Cooperative Plan Execution}

The executor implements a cooperative plan by assigning its roles to available CAVs and applying the specified speed controls at the prescribed times. For a selected plan $\theta$, vehicles are chosen from the trigger zone $\mathcal Z$ according to their lane roles $r_i$ and their front-to-back order.

After assignment, each selected vehicle $i$ waits for $\tau_i$ before its speed-control action begins. During this waiting period, the vehicle remains in its lane and follows the baseline policy $\pi_0$ for longitudinal control. Once the activation delay has elapsed, the executor controls the vehicle toward the target speed $v_i^{\mathrm h}$ for a planned duration $T_i$. The vehicle is released earlier if it reaches the lane drop before the holding period ends.

In speed control, the executor computes an acceleration request from the difference between target and current speeds:
\begin{equation}
a_{i,t}^{\theta}
=
\operatorname{clip}
\left(
\frac{v_i^{\mathrm h}-v_{i,t}}{\Delta t_c},
-d_i,a_{\max}
\right),
\label{eq:hold_control}
\end{equation}
where $v_{i,t}$ is the current speed, $\Delta t_c$ is the control interval, $d_i$ is the maximum allowed deceleration, and $a_{\max}$ is the maximum allowed acceleration. The clipping operation keeps the requested acceleration within these limits.

The executor switches between the plan and $\pi_0$ based on whether the vehicle is in its speed-control period. Let $a_{i,t}^{0}$ denote the acceleration requested by $\pi_0$, and let $q_{i,t}=1$ when the plan controls the vehicle's longitudinal motion and $q_{i,t}=0$ otherwise. The resulting acceleration request is
\begin{equation}
a_{i,t}^{\mathrm{req}}
=
q_{i,t}a_{i,t}^{\theta}
+
(1-q_{i,t})a_{i,t}^{0}.
\label{eq:control_switch}
\end{equation}
Thus, the baseline request is used before speed control begins and after the vehicle is released. The executor applies safety limits to the request and sends SUMO a speed target for the next control interval. For subsequent evaluation, the controller that uses only plan $\theta$ is denoted by $\pi_{\{\theta\}}$.

\subsection{Physics-Informed Traffic World Model}
\label{sec:world_model}

Searching for cooperative plans requires evaluating many combinations of vehicle roles, target speeds, and execution times. Directly testing every candidate in SUMO across multiple traffic realizations can be computationally expensive. We therefore use a traffic world model to predict the evolution of mixed traffic under candidate plans and guide the selection of plans for subsequent SUMO verification.

The world model combines traffic rules with learned corrections. For a human-driven vehicle (HDV) $v$, let $h_{v,t}$ denote its observed motion history at time $t$, and let $\phi$ denote the learned model parameters. An encoder estimates the vehicle's car-following and lane-changing parameters:
\begin{equation}
\hat{\eta}_{v,t}
=
E_\phi(h_{v,t}).
\label{eq:driver_encoding}
\end{equation}
Given these parameters and the current traffic state, the intelligent driver model (IDM) computes a base acceleration, while lane-changing rules assign a score to each candidate lane. Learned corrections adjust the acceleration and lane scores separately, with gates controlling the magnitude of the adjustments. The corrected outputs are used to update vehicle speeds, positions, and lanes over successive prediction steps. The learned components are trained on recorded vehicle trajectories.

To evaluate a candidate plan, the world model predicts traffic evolution with and without the plan from the same initial state. In the plan prediction, CAVs are selected according to their assigned roles. After the specified activation delays, their longitudinal motion is controlled toward the target speeds for the corresponding holding durations. Changes in the controlled CAVs' speeds and positions affect the predicted motion of surrounding vehicles, and these interactions propagate over successive prediction steps. In the no-plan prediction, vehicles follow the world model's own motion rules. The difference between the two predicted road delays estimates the effect of the candidate plan.

Let $\widehat L_\phi(\theta,s_j,\omega_{jk})$ denote the predicted cumulative delay on the road when plan $\theta$ is applied from state $s_j$ with random inputs $\omega_{jk}$. Using $\varnothing$ to denote no plan, the mean paired delay saving at state $s_j$ is
\begin{equation}
\bar r_{\theta,j}
=
\frac{1}{K}\sum_{k=1}^{K}
\left[
\widehat L_\phi(\varnothing,s_j,\omega_{jk})
-
\widehat L_\phi(\theta,s_j,\omega_{jk})
\right].
\label{eq:predicted_saving}
\end{equation}
A positive value indicates that the plan is predicted to reduce road delay relative to the model's no-plan prediction. Both predictions continue to the configured horizon, even if the plan ends earlier, so that the comparison also accounts for predicted traffic effects after the intervention.

The no-plan prediction follows the world model's own motion rules and does not use MAPPO actions. In addition, the predicted delay excludes entrance waiting, whereas the SUMO objective $J$ includes it. Therefore, the predicted savings are used to guide candidate selection rather than to measure improvement over MAPPO directly. Shortlisted plans are subsequently evaluated against MAPPO in SUMO.

\subsection{World-Model-Guided Plan Search}
\label{sec:plan_search}

Evolutionary search explores cooperative plans, starting from manually designed and randomly generated candidates. Each generation retains high-scoring plans and creates new ones by modifying their speed-control parameters and execution timing.

For a plan $\theta$, let $r_{\theta,jk} = \widehat L_\phi(\varnothing,s_j,\omega_{jk}) - \widehat L_\phi(\theta,s_j,\omega_{jk})$ be the predicted delay saving in the $k$-th paired rollout from state $s_j$. The mean savings across the $n$ starting states form $\mathcal R_\theta=\{\bar r_{\theta,j}\}_{j=1}^{n}$. We score the plan as
\begin{equation}
\begin{aligned}
d_\theta
&=
\frac{1}{n}\sum_{j=1}^{n}
\operatorname{SD}
\left(
\{r_{\theta,jk}\}_{k=1}^{K}
\right),\\
S_\phi(\theta)
&=
\operatorname{LCB}_{0.95}(\mathcal R_\theta)
-
\lambda\frac{d_\theta}{\sqrt{n}},
\end{aligned}
\label{eq:search_score}
\end{equation}
where $d_\theta$ measures the average variation across repeated predictions from the same state, and $\lambda\geq 0$ controls the penalty. The lower confidence bound favors plans with higher estimated savings across starting states, while the penalty discourages plans whose predictions vary substantially under repeated rollouts. The confidence bound is computed by bootstrap resampling of traffic scenarios, keeping starting states from the same scenario together.

After removing infeasible candidates, duplicates, and plans reserved for comparison, the remaining candidates form $\Theta_{\mathrm{cand}}$. We select the highest-scoring candidates for simulator verification:
\begin{equation}
\mathcal C_{\mathrm{ver}}
=
\operatorname{Top}_{B}
\left(
\Theta_{\mathrm{cand}},S_\phi
\right),
\label{eq:candidate_selection}
\end{equation}
where $\operatorname{Top}_{B}$ selects up to $B$ distinct plans. Thus, world-model predictions guide both candidate updates and the allocation of the limited SUMO verification budget.

\subsection{Simulator Verification and Plan Library}
\label{sec:verification}

The world-model score ranks candidates, but it does not directly measure their improvement over MAPPO. We therefore compare each selected plan with the baseline policy $\pi_0$ in SUMO under matched traffic realizations. For realization $\xi$, the total delay difference is
\begin{equation}
\Delta J_\theta(\xi)
=
J(\pi_{\{\theta\}},\xi)
-
J(\pi_0,\xi).
\label{eq:delay_difference}
\end{equation}
A negative value indicates that the plan reduces total delay relative to MAPPO in that realization.

A reduction in total delay may still be accompanied by additional delay for particular vehicle groups. We therefore admit a plan to the library only if it satisfies both a total-delay criterion and a group-delay criterion:
\begin{equation}
\begin{aligned}
\operatorname{UCI}_{0.95}
\left(
\{\Delta J_\theta(\xi)\}_{\xi\in\Xi_{\mathrm{ver}}}
\right)
&<0,\\
\frac{1}{|\Xi_{\mathrm{ver}}|}
\sum_{\xi\in\Xi_{\mathrm{ver}}}
\Delta D_g(\{\theta\},\xi)
&\leq\epsilon,
\quad \forall g\in\mathcal G,
\end{aligned}
\label{eq:plan_admission}
\end{equation}
where $\operatorname{UCI}_{0.95}$ is the upper endpoint of the two-sided 95\% confidence interval for the mean paired delay difference. The first condition requires evidence of lower mean total delay than MAPPO; the second limits the mean additional delay for each vehicle group across the verification realizations. We consider CAVs, HDVs, and vehicles originating from each of the three lanes, with $\epsilon=10\,\mathrm{s}$ per vehicle.

The retained plans form the library $\mathcal B$ and are ordered by the lower confidence bounds of their verified delay savings. At runtime, the executor checks trigger conditions, cooldown requirements, and vehicle availability before activating the first feasible plan. If a selected vehicle is already assigned to another plan, the activation attempt is skipped. The complete controller $\pi_{\mathcal B}$ is evaluated separately to assess the combined effects of the retained plans.

Together, world-model-guided search concentrates simulator verification on promising candidates, while SUMO screening checks their benefits and group-level costs against MAPPO. This combination supports plan discovery under a limited verification budget and provides a reusable library for cooperative control.

\begin{table*}[!ht]
    \centering
    \caption{Traffic delay across 192 paired lane-drop scenarios at 60\% CAV penetration and 2,800 vehicles/h. Brackets give 95\% bootstrap confidence intervals.}
    \label{tab:lane-drop-main-results}

    \small
    \setlength{\tabcolsep}{4pt}
    \renewcommand{\arraystretch}{1.19}

    \begin{tabularx}{\textwidth}{>{\raggedright\arraybackslash}Xcccc}
        \toprule
        & & \multicolumn{3}{c}{Change relative to MAPPO} \\
        \cmidrule(l){3-5}
        Method
            & Mean total delay
            & Total delay
            & Most-delayed 10\%
            & Largest group increase \\
        & (veh\,s/scenario) & (\%) & (s/vehicle) & (s/vehicle) \\
        \midrule

        MAPPO (reference)
            & $34{,}746\;[33{,}192,\,36{,}351]$
            & $0$
            & $0$
            & $0$ \\

        Two-CAV hold
            & $30{,}896\;[29{,}782,\,32{,}056]$
            & $-11.1\;[-13.8,-8.3]$
            & $-33.8\;[-39.7,-28.1]$
            & $\mathbf{+4.1}\;[3.6,4.6]$ \\

        Hand-written library
            & $30{,}977\;[29{,}881,\,32{,}092]$
            & $-10.8\;[-13.8,-7.7]$
            & $-42.4\;[-48.5,-36.4]$
            & $+7.1\;[6.6,7.7]$ \\

        \rowcolor{LaneDropTint}
        \textbf{Searched library}
            & $\mathbf{29{,}790}\;[28{,}831,\,30{,}784]$
            & $\mathbf{-14.3}\;[-16.9,-11.5]$
            & $\mathbf{-47.2}\;[-53.0,-41.5]$
            & $+6.2\;[5.7,6.7]$ \\
        \bottomrule
    \end{tabularx}

    \par\smallskip
    \parbox{\textwidth}{\footnotesize
        Bold values mark the lowest point estimate among the three interventions in each column. MAPPO's changes are zero by definition. For all three interventions, the largest group increase occurs in origin lane L2.}
\end{table*}

\section{Experiments}
\label{sec:lane-drop-experiments}

\subsection{Experimental Setup}
\label{sec:experimental_setup}

\subsubsection{Scenario and Baselines}

Experiments are conducted in SUMO on a three-to-two-lane drop scenario with a traffic demand of 2,800 vehicles/h and a CAV penetration rate of 60\%. Four methods are compared: MAPPO alone, a fixed two-CAV speed hold, a hand-written plan library, and a plan library obtained through world-model-guided search. All methods use the same frozen MAPPO policy trained at the evaluated penetration rate. The two plan libraries also use the same executor.

\subsubsection{Plan Screening}

Candidate plans are screened against MAPPO using 24 shared traffic scenarios. A plan passes the screening if the upper bound of the 95\% confidence interval for its paired total-delay change is below zero and the mean delay increase does not exceed 10\,s per vehicle in any of the five groups: vehicles from each of the three origin lanes, CAVs, and HDVs. For each maneuver form, the passing plan with the lowest upper confidence bound is retained. The resulting libraries retain one of the ten manually designed plans and three of the twelve shortlisted search candidates, respectively.

\subsubsection{Evaluation Protocol}

The four methods are evaluated on 192 additional scenarios, paired across methods by traffic seed. Total delay includes both entrance waiting and delay accumulated while traveling below 31.29\,m/s. Among vehicles present in both runs, tail delay is measured as the mean delay of the most-delayed 10\% under each method.

Percentile 95\% confidence intervals are obtained from 10,000 bootstrap resamples, with scenarios resampled jointly across methods. Percentage changes are calculated using the reference method's mean total delay as the denominator. One of two search configurations is selected according to a rule fixed before testing. The primary exploratory comparison is between the searched and hand-written plan libraries.

\subsection{Traffic Performance of the Verified Library}

The traffic performance of the searched plan library is compared with MAPPO, a fixed two-CAV hold, and a hand-written plan library. The fixed hold assigns the same speed target simultaneously to one CAV in each through lane. The hand-written library retains one single-CAV hold from ten manually designed maneuvers, whereas the searched library contains three plans with two to four CAV roles. All methods use the same MAPPO policy, and both libraries are subject to the same SUMO screening criteria and executor.

Across 192 paired test scenarios, the searched library reduces mean total delay by 14.3\% relative to MAPPO and by 3.6\% relative to the fixed two-CAV hold (Table~\ref{tab:lane-drop-main-results}). It also reduces total delay by 3.8\% relative to the hand-written library, saving 1,186 vehicle-seconds per scenario (95\% CI: $[382,\,2{,}000]$). Thus, the benefit extends beyond adding a fixed cooperative intervention to MAPPO: the searched library also improves on the tested hand-written library under the same screening and execution rules.

The reduction in total delay is accompanied by lower tail delay. The mean delay of the most-delayed 10\% of vehicles is 47.2\,s per vehicle lower than under MAPPO and 4.8\,s lower than under the hand-written library. This result indicates that the aggregate improvement is accompanied by a reduction in measured tail delay.

The results also reveal a trade-off across vehicle groups. Vehicles originating in lane~L2 experience the largest mean delay increase relative to MAPPO: 6.2\,s per vehicle under the searched library, compared with 4.1\,s under the fixed hold. Although the searched library achieves lower total and tail delay, it imposes a larger mean cost on this group than the fixed hold. The increase remains below the 10\,s-per-vehicle allowance, as do the increases for all other evaluated groups.

\begin{table}[!ht]
\centering
\caption{Candidate yield under the same 24-scenario SUMO screen.}
\label{tab:lane-drop-discovery}
{\footnotesize
\renewcommand{\arraystretch}{1.12}
\setlength{\tabcolsep}{3pt}
\begin{tabular*}{\columnwidth}{@{\extracolsep{\fill}}lrrr@{}}
\toprule
Candidate source & Screened & Passed & Pass rate\\
\midrule
World-model-guided search & 500 & 150 & 30.0\%\\
Random draws & 1,500 & 39 & 2.6\%\\
\bottomrule
\end{tabular*}}
\end{table}

\subsection{Candidate Discovery with World-Model Guidance}

Random search measures how often a plan drawn from the feasible plan space passes SUMO verification. We use this baseline to test whether the world model can identify promising combinations of vehicle roles and execution times before sending them to SUMO. Discovery efficiency is measured by the fraction of screened candidates that meet both the total-delay and vehicle-group cost criteria.

As shown in Table~\ref{tab:lane-drop-discovery}, world-model-guided search found 150 qualifying plans, compared with 39 for random search, despite sending only one-third as many candidates to SUMO. The observed pass rate was 30.0\% with world-model guidance and 2.6\% with random search. At these rates, SUMO screened an average of 3.3 candidates per qualifying plan with guidance, compared with 38.5 under random search. This advantage matters when simulation capacity is limited: the world model directs testing toward promising combinations of vehicle roles and execution times, and SUMO verifies which combinations reduce total delay while keeping the cost to each vehicle group within the prescribed limit.

\begin{figure*}[!ht]
    \centering
    \subfloat[Total delay by role count\label{fig:role-count}]{%
        \includegraphics[width=0.48\textwidth]{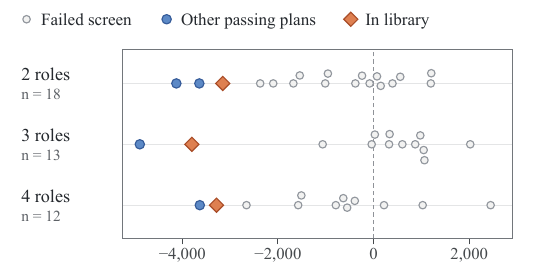}}%
    \hfill
    \subfloat[Two-CAV control schedules\label{fig:two-cav-timing}]{%
        \includegraphics[width=0.48\textwidth]{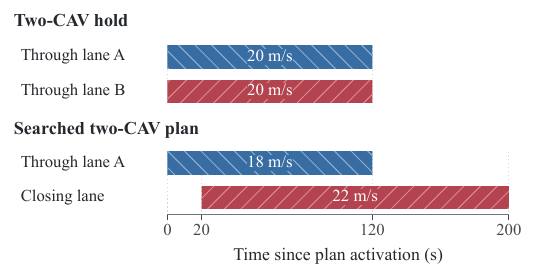}}
    \caption{(a) Each point shows one plan's mean change in total delay relative to MAPPO across 24 paired scenarios (vehicle-seconds per scenario). Negative values indicate savings. (b) Bars show scheduled control windows and target speeds for the fixed and searched two-CAV plans.}
    \label{fig:lane-drop-structure}
\end{figure*}

\subsection{Vehicle Roles and Execution Timing}

To examine how much the number of participating CAVs explains the observed delay savings, we compare 43 candidate plans on the same 24 SUMO scenarios, grouping them by the number of assigned roles. In Fig.~\ref{fig:lane-drop-structure}(a), all three groups contain plans on both sides of the MAPPO reference. Among the 18 two-CAV plans, for example, mean total delay ranged from 4,124 vehicle-seconds below MAPPO to 1,206 vehicle-seconds above it per scenario. The two-CAV plan retained in the library, marked by the diamond in the first row, reduced mean total delay by 3,153 vehicle-seconds per scenario and achieved lower mean total delay than 10 of the 12 four-CAV plans. Thus, even among plans evaluated under identical traffic conditions, assigning more vehicles did not consistently yield greater savings.

Fig.~\ref{fig:lane-drop-structure}(b) shows how the retained two-CAV plan coordinates its participants. The fixed hold assigns one CAV to each through lane and schedules both to target 20\,m/s for the first 120\,s. The searched plan gives the two vehicles distinct roles: the through-lane CAV targets 18\,m/s from 0 to 120\,s, and the closing-lane CAV targets 22\,m/s from 20 to 200\,s. This schedule creates 100\,s of overlap, followed by 80\,s during which only the closing-lane CAV is scheduled to act. With the same two-vehicle allocation, the search therefore changes both the lanes involved and the sequence of control, identifying a plan that uses coordinated action across the through and closing lanes.

\subsection{Delay Savings over Time and Bottleneck Throughput}

\begin{figure}[!t]
    \centering
    \includegraphics[width=\columnwidth]{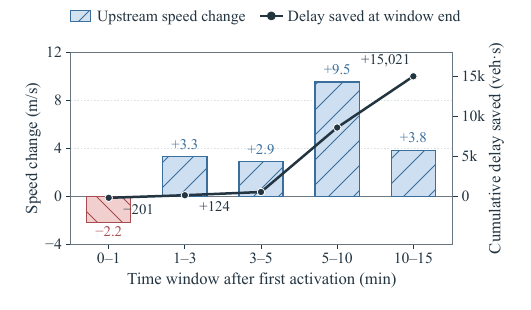}
    \caption{One paired scenario relative to MAPPO. Bars show the mean upstream speed change within each window. The line shows cumulative delay saved by each window's end.}
    \label{fig:lane-drop-time}
\end{figure}

Fig.~\ref{fig:lane-drop-time} follows one paired scenario for 15 minutes after the first plan activation. During the first minute, mean speed in the 500\,m upstream of the bottleneck was 2.2\,m/s lower under the searched library than under MAPPO, and the library accumulated 201 additional vehicle-seconds of delay. By the end of minute~3, the library had recovered this initial cost and saved 124 vehicle-seconds. The largest upstream speed advantage, 9.5\,m/s, occurred during minutes~5--10, as cumulative savings grew from 526 to 8,602 vehicle-seconds. Savings reached 15,021 vehicle-seconds by minute~15. In this scenario, evaluating the plan after one minute would have reversed the ranking observed over the full evaluation window.

As shown in Table~\ref{tab:delay-throughput}, across the 192 paired scenarios, the searched library reduced total delay in 147 cases and increased the number of vehicles crossing the bottleneck in 155 cases over the same 15-minute window. Both improvements occurred together in 131 scenarios (68.2\%). The remaining cases show why the two measures are worth reporting together. In 24 scenarios, more vehicles crossed the bottleneck even though total delay increased. Conversely, 16 scenarios recorded lower delay with the same or fewer crossings. Thus, higher throughput accompanied most of the observed delay reductions, while the paired comparison also identifies cases in which the two outcomes diverged.

\begin{table}[!t]
    \centering
    \caption{Paired scenario counts by 15-minute outcome relative to MAPPO.}
    \label{tab:delay-throughput}
    \footnotesize
    \begin{tabular}{@{}lrrrr@{}}
        \toprule
        & \multicolumn{3}{c}{Vehicles crossing the bottleneck} & \\
        \cmidrule(lr){2-4}
        Total delay & Fewer & Same & More & Total \\
        \midrule
        Lower  & 12 & 4 & \textbf{131} & 147 \\
        Higher & 20 & 1 & 24           & 45  \\
        \midrule
        Total  & 32 & 5 & 155          & 192 \\
        \bottomrule
    \end{tabular}
\end{table}

\section{Conclusion}
\label{sec:conclusion}

This work presented an offline framework integrating world-model guidance with simulator verification to discover cooperative control plans at lane-drop bottlenecks. Joint optimization of vehicle roles and timing produced a reusable plan library that achieved lower mean total delay than the tested manually designed controls while satisfying the empirical limit on additional group mean delay. Guided search also achieved a higher candidate-screening pass rate than random sampling under the same acceptance criteria. Further analysis showed that increasing vehicle participation does not consistently improve performance and illustrated how initial holding costs can be offset by later traffic benefits. The current evaluation is limited to one road layout, demand level, and CAV penetration rate. Future work will evaluate generalization across bottleneck geometries, traffic demands, and automated vehicle penetration rates, while assessing robustness to variations in traffic conditions.

\bibliographystyle{IEEEtran}
\bibliography{refs}

\end{document}